\documentclass[10pt,letterpaper]{article}
\usepackage[margin=1in]{geometry}
\usepackage{graphicx}
\usepackage{amsmath}
\usepackage{xcolor}
\usepackage{hyperref}
\usepackage{setspace}
\begin{document}
	
\title{Strict Enforcement of Conservation Laws and Invertibility in CNN-Based Super Resolution for Scientific Datasets}

\author{Andrew Geiss$^{*1}$ and Joseph C. Hardin$^{1,2}$\\
	1 - Pacific Northwest National Laboratory\\
	2 - Climate AI\\
	{\tt\small andrew.geiss@pnnl.gov, josephhardinee@gmail.com}\\
	{\tiny \textit{$^*$- Corresponding Author}}
}

\maketitle

\begin{abstract}
Recently, deep Convolutional Neural Networks (CNNs) have revolutionized image super-resolution (SR), dramatically outperforming past methods for enhancing image resolution. They could be a boon for the many scientific fields that involve image or gridded datasets: satellite remote sensing, radar meteorology, medical imaging, numerical modeling etc. Unfortunately, while SR-CNNs produce visually compelling outputs, they may break physical conservation laws when applied to scientific datasets. Here, a method for ``Downsampling Enforcement" in SR-CNNs is proposed. A differentiable operator is derived that, when applied as the final transfer function of a CNN, ensures the high resolution outputs \textit{exactly} reproduce the low resolution inputs under 2D-average downsampling while improving performance of the SR schemes. The method is demonstrated across seven modern CNN-based SR schemes on several benchmark image datasets, and applications to weather radar, satellite imager, and climate model data are also shown. The approach improves training time and performance while ensuring physical consistency between the super-resolved and low resolution data.
\end{abstract}

\pagebreak

\section{Introduction}
Image Super-Resolution (SR) involves increasing the resolution of images beyond their native resolution and is a long-standing problem in the field of image processing. Here, we focus on Single Image Super Resolution (SISR) which involves estimating sub-pixel scale values based only on a single coarsely resolved input image \cite{sr_review} (as opposed to SR frameworks that utilize multiple images \cite{multi_view, multi_view_misr,video_sr,video_sr2}). The simplest approach to SISR is 2D-interpolation, and many schemes exist that produce High Resolution (HR) outputs of various qualities. Sophisticated SISR schemes can perform different operations at different locations in an image depending on the local Low Resolution (LR) pixel data: using a dictionary of image-patch exemplars for instance \cite{a_plus}. Nasrollahi and Moeslund (2014) \cite{sr_review} provide an overview of SISR schemes.

\subsection{Super Resolution with Neural Networks}
Recently, deep Convolutional Neural Networks (CNNs) have been applied to SISR and have significantly outperformed past algorithms. In 2016, a 3-layer CNN achieved state of the art SISR results \cite{srcnn}, and the approach was quickly expanded to use significantly deeper CNNs \cite{deep_srcnn}. SISR CNN architectures have rapidly developed, and complex SISR networks are now built up of many blocks of convolutional layers that include skip connections such as dense blocks \cite{densenet}, residual blocks \cite{resnet}, and  channel attention blocks \cite{can}. The CNNs' internal spatial upsampling operators have progressed from bicubic upsampling \cite{srcnn}, to learned kernels \cite{fully_conv}, to the ``pixel-shuffle'' approach \cite{pixel_shuffle}. The loss functions have also evolved from simple pixel-wise errors, to feature-loss and adversarial-loss\cite{srgan} which allow the CNNs to hallucinate plausible sub-pixel scale features. Current CNN SR schemes combine many of these concepts \cite{drln}. Wang, Chan and Hoi (2020) \cite{cnn_sr_review} review CNN-based SISR.

\subsection{Invertible Super Resolution Networks}
SISR CNNs are not typically invertible. Training them usually involves degrading HR images and tasking the CNN to reconstruct them, but applying the same degradation to the CNN output does not necessarily reproduce the input image. SISR is an ill-posed problem because there are usually multiple HR images that produce the same LR image when downsampled, and constraining SISR CNN outputs to this manifold of possible HR images is desirable \cite{pulse}. Several studies have approximated this type of invertibility with a modified loss function that computes the pixel-error between the downsampled output and the LR input \cite{pulse,dsloss1,dsloss2,zhang_2020}. Additionally, in cases where the HR image is known but needs to be intentionally degraded (image compression) training two CNNs simultaneously to perform upsampling and downsampling provides better performance than other SR schemes \cite{learned_downscaling, kim_2018}. These approaches only approximate invertibility however, and may not be sufficient for super-resolving scientific datasets.

\subsection{Contributions and Impacts}
This study introduces a method referred to as ``Downsampling Enforcement'' (DE), that strictly constrains a super resolution CNN's output to be \textit{exactly} invertible under 2D-average downsampling. This is accomplished using transfer function that is applied after the last convolutional layer of the CNN. We demonstrate this method using seven different CNN SISR architectures on five common image datasets, and find that it improves performance in every case. We also demonstrate the method on scientific datasets from a weather radar, satellite imager, and climate model; all cases where strictly enforcing conservation laws is important.

\subsection{Motivation} \label{applications_sec}
There are many scientific fields where CNN-based SR could be applied to image data or gridded datasets. In these applications, guaranteed physical consistency under 2D-averaging can ensure the SR scheme obeys physical conservation laws. For example, satellite imagers often have a much larger dynamic range than handheld cameras and undergo rigorous calibration and validation to ensure that measured radiances are accurate \cite{landsat_calib}; this should be considered when applying CNN-based super-resolution \cite{liebel_2016, lanaras_2018, muller_2020}. Other possible applications of this method include data from ranging instruments such as radars \cite{radarsr}, sonars \cite{sonar_sr}, and lidars \cite{lidar_sr}. Weather radars can be used to estimate precipitation rates for instance \cite{nexrad_rainfall}, a physical quantity that should be conserved under spatial averaging. Super resolution can be used to enhance output from gridded numerical simulations \cite{downscaling_sr}, and CNN-SISR has already been demonstrated on several real-world numerical simulation problems, including: precipitation modeling \cite{wang_2021}, wind and solar modeling \cite{stengel_2020}, and climate modeling \cite{vandal_2018}. In climate simulations strict enforcement of conservation laws is of particular importance because climate signals can be relatively weak. Also, if downscaling is used during model integration, even small errors can grow rapidly over many time-steps and significantly impact results. The lack of an internal representation of physics or strict adherence to physical laws in CNNs has been identified as a major hurdle that must be addressed before their impressive capabilities can be fully brought to bear on important imaging and modeling problems in the physical sciences \cite{reichstein_2019, tsagkatakis_2019}. In these SISR applications, and many others, strict conservation of large-scale statistical properties is often just as important as the visual fidelity of the HR output, and our method can ensure both.

\section{The Downsampling Enforcement Operator} \label{f_interp_sec}
Typically, during training, CNN-SISR schemes are provided LR input images produced by degrading HR-images and tasked with recovering the original. If $I_{HR}$ and $I_{LR}$ are the high- and low-resolution images respectively, $D$ is the image downsampling operator and $S$ is the super resolution scheme, CNN-SISR schemes try to find $S$ such that: $I_{HR} \approx S\{I_{LR}\}$, and during training: $I_{HR} \approx S\{D\{I_{HR}\}\}$. Here, we derive a ``Downsampling Enforcement" (DE) operator, that can be incorporated into most common SR CNN and ensures the CNN also satisfies: $I_{LR} = D\{S\{I_{LR}\}\}$. Here, we assume that $D$ represents 2D-average downsampling though solutions can likely be derived for other downsampling schemes. The DE operator $f(\mathbf{x},P)$ operates on each $N\times N$-pixel block in the HR image. $P$ denotes the value of a single pixel in the LR image and $x_i \in \mathbf{x}$ are the $N \times N$ corresponding HR-image pixels output by the last conv-layer in the CNN. $P$ and $x_i$ are assumed to have pixel intensities bounded by $[-1,1]$.

\begin{equation}
f(\mathbf{x},P)_i = \begin{cases}
x_i + \left(\frac{P-\bar{x}}{1-\bar{x}}\right)(1-x_i) & \bar{x} < P \\
x_i & \bar{x} = P \\
x_i + \left(\frac{P-\bar{x}}{1+\bar{x}}\right)(1 + x_i) & \bar{x} > P\\
\end{cases}
\qquad \text{where:} \qquad \bar{x} = \frac{1}{N^2}\sum_{x_j \in \mathbf{x}} x_j
\label{f_piecewise}
\end{equation}

Which can also be written:

\begin{equation}
f(\mathbf{x},P)_i = x_i + (P-\bar{x})\left(\frac{\sigma + x_i}{\sigma + \bar{x}}\right) \qquad \text{where:} \qquad \sigma = \text{sign}(\bar{x}-P) \label{f_compact}
\end{equation}

A detailed derivation of $f(\mathbf{x},P)$ is provided in the Section 1 of the supplement. This formulation of $f$ has several useful properties:

\begin{equation}\frac{1}{N^2}\sum_{i = 0}^{N^2} f(\mathbf{x},P)_i = P \label{summation_constraint} \end{equation}
\begin{equation}f(\mathbf{x},P)_i \in [-1,1] \label{range_constraint} \end{equation}
\begin{equation}x_i > x_j \rightarrow f(\mathbf{x},P)_i \geq f(\mathbf{x},P)_j \label{inequality_constraint}\end{equation}
\begin{equation}f(\mathbf{x},P)_i \text{ is piecewise differentiable} \label{differentiation_constraint}\end{equation}

(\ref{summation_constraint}) ensures invertibility under 2D-average downsampling. (\ref{range_constraint}) bounds $f(\mathbf{x},P)$ to the input image's dynamic range of $[-1,1]$. (\ref{inequality_constraint}) maintains the order of the initial output pixels' intensities. Finally, (\ref{differentiation_constraint}): $f(\mathbf{x},P)$ will be included as a part of the CNN during training and must be differentiable for backpropagation to work. Short proofs that (\ref{f_piecewise}) satisfies these conditions are given in Supplement Section 2.

\begin{figure}[t]
	\begin{center}
		\includegraphics[width=0.5\linewidth]{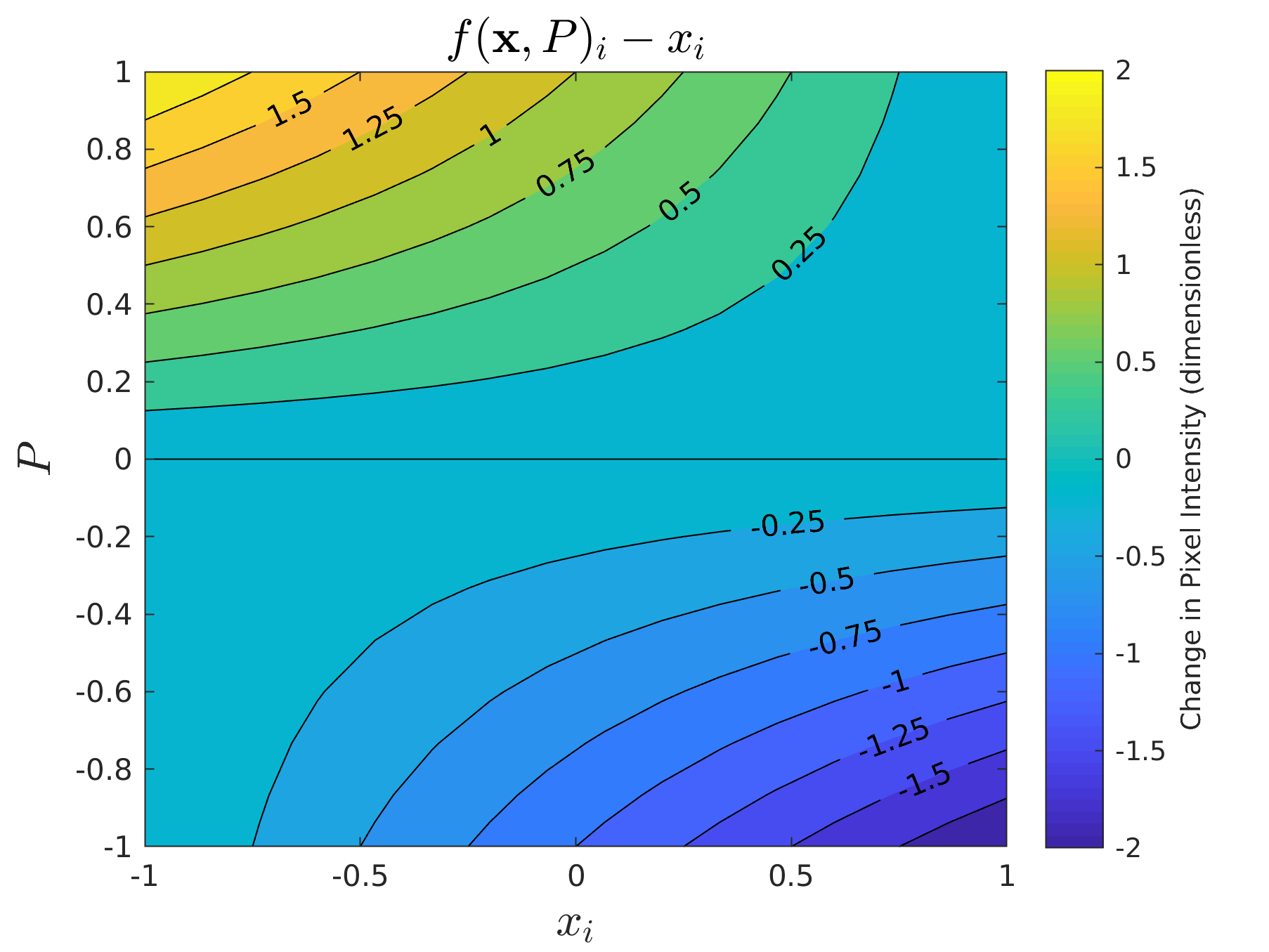}
	\end{center}
	\caption{Visualization of the correction $f(\mathbf{x},P)_i-x_i$ as a function of varying $P$ with a sample input of 16 $x_i$'s ranging from -1 to 1 with a mean of $0$.}
	\label{fx_fig}
\end{figure}

Equation (\ref{f_compact}) has a physical interpretation: it operates on initial SISR-CNN image outputs (the last conv-layer of the CNN prior to applying (\ref{f_compact}) has 3-channel RGB output and a $tanh$ transfer function). (\ref{f_compact}) is a correction applied independently to each channel that ensures the intensity of each $N\times N$ block of HR output pixels ($\mathbf{x}$) exactly averages to the value of the corresponding LR input pixel $P$. When $P$ exceeds $\bar{x}$, the remaining unused output pixel intensity is computed for each output pixel $(1-x_i)$ and a constant fraction of it is added to the output pixel values. A similar approach is applied when $\bar{x}>P$. Figure \ref{fx_fig} shows the magnitude of the correction ($f(\mathbf{x},P)_i-x_i$) when the DE operator is applied to a hypothetical block of output pixels (ranged between $[-1,1]$ with mean 0) for a range of LR input pixel values. It demonstrates that the correction term varies smoothly with respect to $P$ and $x_i$.

\section{Super Resolution of Images}
To evaluate the ``downsampling enforcement'' approach to SISR, we have implemented a selection of  SISR-CNNs from the recent literature and trained them under identical conditions both with and without the DE operator. This section demonstrates the method on image datasets frequently used in the SR literature. In Section \ref{sds_section} we apply the method to several scientific datasets where strict enforcement of conservation laws more important.

\subsection{Neural Networks} \label{cnns_section}
Here, we have reproduced seven different CNN architectures from the recent SISR literature. For unbiased comparison, we have altered each model slightly so they all have a similar number of trainable parameters: $5 \times 10^6$ (except for SR-CNN and Lap-SRN which have fewer). The CNNs were implemented in Keras with a Tensorflow backend and the code and model diagrams can be found on github\footnote{\url{https://github.com/avgeiss/invertible_sr}}. We provide a more detailed overview of the CNNs and our implementations in the Supplement Section 3. The CNNs are: SR-CNN \cite{srcnn}, Lap-SRN \cite{lapsrn}, Dense U-Net (DUN) \cite{unet,densenet,radarsr}, Deep Back Projection Network (DBPN) \cite{dbpn}, Dense SR Net (DSRN) \cite{densesr}, Enhanced Deep Residual Network (EDRN) \cite{edrn}, and Residual Dense Network (RDN) \cite{rdn}. Each are trained both with and without strictly enforced invertibility.

\subsection{Image Datasets}
The Div2k \cite{div2k} dataset was used for training. It contains 800 high resolution training images with a 100-image test set. The last 10 training images are held out and used to compute validation scores \cite{edrn}. Trained CNNs are evaluated on  several image datasets that were used because of their prevalence in the SISR literature \cite{cnn_sr_review}: SET5 \cite{set5}, SET14 \cite{set14}, BSDS100 \cite{bsds100}, Manga109 \cite{manga109}, Urban 100 \cite{urban100} and the 100-image Div2k validation set \cite{div2k}. Manga 109 are illustrated images and Urban 100 contains photographs of urban scenes while the other datasets contain miscellaneous photographs.

\subsection{Training and Testing} \label{training_section}

This study uses 2D-average downsampling for image degradation. Bicubic downsampling with anti-aliasing is more common in the SISR literature; specifically the Matlab scheme, but 2D-averaging was assumed in deriving (\ref{f_compact}). The CNNs here perform 4x SISR, converting 48x48 pixel inputs to 192x192 outputs. Images are standardized to a [-1,1] scale and a $tanh$ activation is applied to the output. In the DE cases the $tanh$ activation is applied \textit{before} applying (\ref{f_compact}). Each CNN is trained for 300 epochs, with the learning rate reduced by a factor of 10 after the 200th epoch. Epochs are 1000 batches of 16 image chips selected randomly from the training set with random flips and rotations. Pixel-wise mean squared error (MSE) is used as a loss function and the Adam optimizer is used with an initial learning rate of $10^{-4}$, $\beta_1 = 0.9$, $\beta_2 = 0.999$, and $\epsilon=10^{-7}$ \cite{edrn,rdn}.

Two evaluation metrics are used: Peak Signal to Noise Ratio (PSNR) \cite{cnn_sr_review} and the Structural Similarity Index (SSIM) \cite{ssim}. PSNR is computed on the intensity (Y) channel after converting the CNN's output to the YCbCr color space \cite{srcnn}. SSIM is a metric designed to be more representative of the perceptual quality of an image than pixel-wise metrics. It scales from -1 to 1 and higher values are better. During validation and testing, each LR image is broken into 48x48 pixel chips using a 24-pixel stride and PSNR and Structural Similarity Index (SSIM) \cite{ssim} are then calculated on the 96x96 center portions of each of the HR outputs.

\begin{figure*}
	\begin{center}
		\includegraphics[width=1\linewidth]{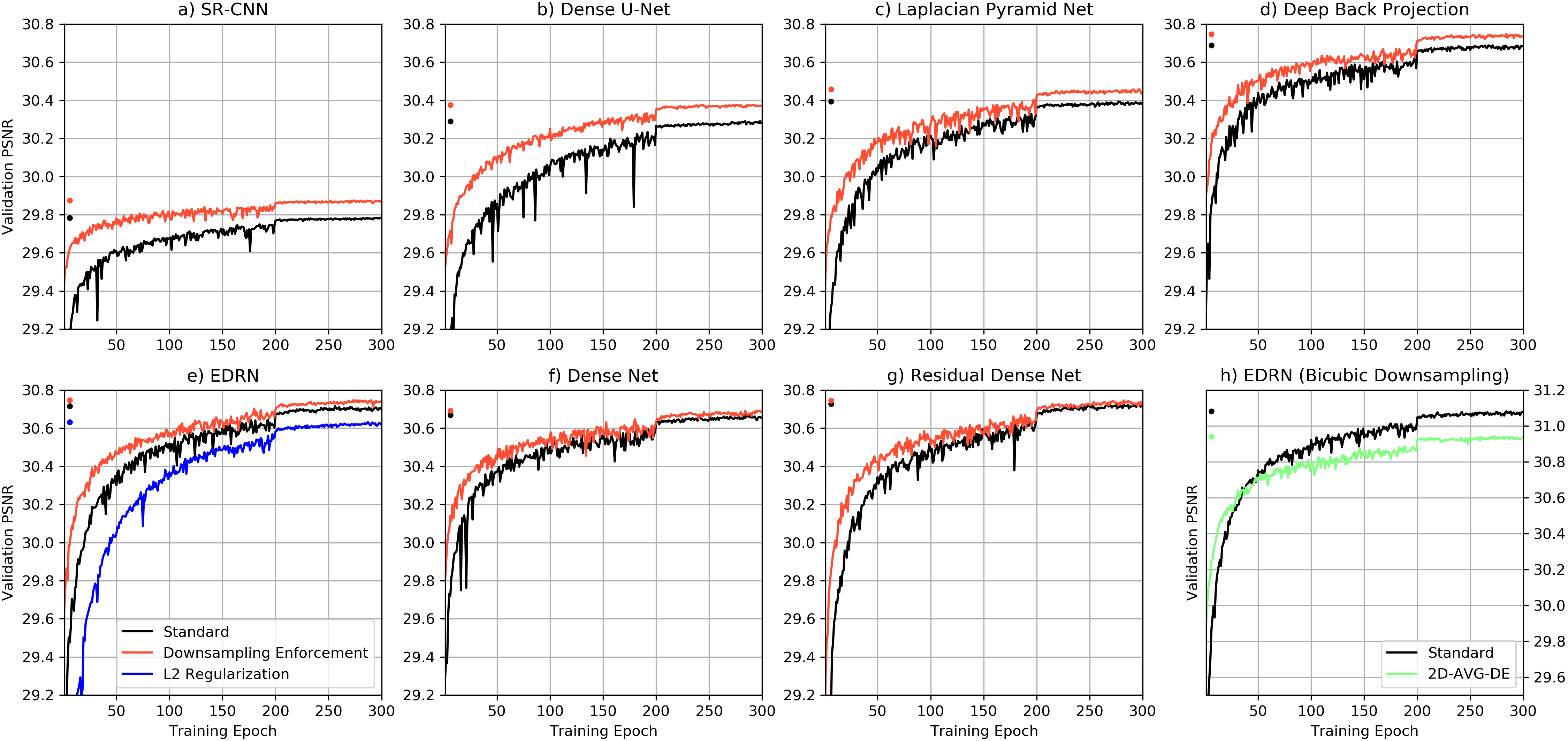}
	\end{center}
	\caption{(a-g): Validation set PSNR during training for the seven CNN architectures with (red) and without (black) Downsampling Enforcement (DE). The blue line in (e) uses an additional loss function term instead of DE. (h): Applies conventional CNN-SISR (black) and DE-SISR that incorrectly assumes 2D-average downsampling (green) to bicubic-downsampled images.}
	\label{training_loss}
\end{figure*}

For each CNN, both with and without DE, a five-member ensemble was trained from randomly initialized weights and the ensemble mean test scores are reported in Table \ref{test_data_table}. Figure \ref{training_loss} shows PSNR computed throughout training on the 10-image validation set for the first ensemble member for each CNN.

\subsection{Results}
Figure \ref{training_loss} (a-g) shows validation PSNR loss during training for each CNN (panels) both with (red lines) and without (black lines) the DE operator. All of the CNN architectures perform comparably or better when DE is added, with the largest advantage early during training.

Examples of sample outputs from each of the CNNs for each of the training sets both with and without DE are included in the supplementary material (Section 4, Figures 1-2). While there are some differences on close inspection, the small differences in PSNR shown in Figure \ref{training_loss} do not relate to any dramatic change in perceptual image quality of the output. These figures help demonstrate that the DE approach can achieve state of the art SISR performance while strictly enforcing physical conservation laws within the CNN architecture.

Table \ref{test_data_table} summarizes final performance for every CNN/test-set pair, with the better scores denoted by \textbf{bold text}. Adding DE to the CNN improved performance in all but one case (the Dense-Net had slightly worse PSNR on Set5, though note that Set5 has only 5-images and results in this column are more likely to be affected by small sample size). The improvements are often small, but are of comparable size to recent generational improvements in CNN architectures, EDRN vs. RDN for instance.\footnote{\url{https://paperswithcode.com/sota/image-super-resolution-on-bsd100-4x-upscaling} Accessed: 28-Jan-2021} Furthermore, in most cases the improvement in the mean test score due to adding DE passes a $99\%$ confidence test (one-sided t-Test for difference in means \cite{rice_2006}). These cases are shown in \textcolor{red}{\textbf{bold red text}}. Overall, the results in Table \ref{test_data_table} show that in addition achieving the primary goal of exact enforcement of conservation rules between the input and output images our approach yields robust and consistent performance improvements when applied across a large sampling of CNN-types and image datasets.

\begin{table*}[t]
	\begin{center}
		\caption{Evaluation of several super resolution CNN architectures, both with and without Downsampling Enforcement (DE), applied to standard test datasets for image super resolution. Entries show Peak Signal to Noise Ratio / Structural Similarity Index (PSNR/SSIM), with higher scores in bold. Values are averaged across 5 training runs with random initializations and red-colored entries pass a 99\% confidence test for difference in means.}
		
		\begin{small}
			\begin{tabular}{ c || c | c | c | c | c | c }
				& SET5 & SET14 & BSDS100 & Manga-109 & Urban-100 & Div2k \\ \hline
				SR-CNN & 32.26/0.8914 & 27.06/0.7466 & 26.34/0.7151 & 27.40/0.8444 & 24.22/0.7242 & 29.04/0.8087 \\
				w/ DE & \textcolor{red}{\textbf{32.44}}/\textcolor{red}{\textbf{0.8954}} & \textcolor{red}{\textbf{27.18}}/\textcolor{red}{\textbf{0.7507}} & \textcolor{red}{\textbf{26.41}}/\textcolor{red}{\textbf{0.7181}} & \textcolor{red}{\textbf{27.65}}/\textcolor{red}{\textbf{0.8518}} & \textcolor{red}{\textbf{24.37}}/\textcolor{red}{\textbf{0.7321}} & \textcolor{red}{\textbf{29.15}}/\textcolor{red}{\textbf{0.8134}} \\
				\hline 
				DUN & 33.26/0.9032 & 27.61/0.7593 & 26.67/0.7270 & 28.65/0.8673 & 25.00/0.7566 & 29.57/0.8224 \\
				w/ DE & \textbf{33.30}/\textcolor{red}{\textbf{0.9047}} & \textcolor{red}{\textbf{27.68}}/\textcolor{red}{\textbf{0.7618}} & \textcolor{red}{\textbf{26.72}}/\textcolor{red}{\textbf{0.7290}} & \textcolor{red}{\textbf{28.85}}/\textcolor{red}{\textbf{0.8721}} & \textcolor{red}{\textbf{25.15}}/\textcolor{red}{\textbf{0.7632}} & \textcolor{red}{\textbf{29.66}}/\textcolor{red}{\textbf{0.8256}} \\
				\hline 
				Lap-SRN & 33.22/0.9037 & 27.63/0.7605 & 26.70/0.7289 & 28.86/0.8709 & 25.13/0.7608 & 29.66/0.8249 \\
				w/ DE & \textbf{33.29}/\textcolor{red}{\textbf{0.9048}} & \textcolor{red}{\textbf{27.69}}/\textcolor{red}{\textbf{0.7615}} & \textcolor{red}{\textbf{26.76}}/\textcolor{red}{\textbf{0.7302}} & \textcolor{red}{\textbf{28.94}}/\textcolor{red}{\textbf{0.8720}} & \textcolor{red}{\textbf{25.31}}/\textcolor{red}{\textbf{0.7672}} & \textcolor{red}{\textbf{29.73}}/\textcolor{red}{\textbf{0.8268}} \\
				\hline 
				DBPN & 33.49/0.9075 & 27.85/0.7673 & 26.88/0.7354 & 29.58/0.8824 & 25.72/0.7816 & 29.99/0.8333 \\
				w/ DE & \textbf{33.54}/\textcolor{red}{\textbf{0.9085}} & \textcolor{red}{\textbf{27.88}}/\textcolor{red}{\textbf{0.7688}} & \textcolor{red}{\textbf{26.91}}/\textcolor{red}{\textbf{0.7368}} & \textcolor{red}{\textbf{29.70}}/\textcolor{red}{\textbf{0.8855}} & \textcolor{red}{\textbf{25.86}}/\textcolor{red}{\textbf{0.7871}} & \textcolor{red}{\textbf{30.06}}/\textcolor{red}{\textbf{0.8353}} \\
				\hline 
				EDRN & 33.52/0.9075 & 27.90/0.7676 & 26.89/0.7362 & 29.61/0.8824 & 25.80/0.7853 & 30.04/0.8348 \\
				w/ DE & \textbf{33.58}/\textcolor{red}{\textbf{0.9086}} & \textbf{27.93}/\textcolor{red}{\textbf{0.7690}} & \textcolor{red}{\textbf{26.92}}/\textcolor{red}{\textbf{0.7372}} & \textcolor{red}{\textbf{29.68}}/\textcolor{red}{\textbf{0.8852}} & \textcolor{red}{\textbf{25.88}}/\textcolor{red}{\textbf{0.7882}} & \textcolor{red}{\textbf{30.07}}/\textcolor{red}{\textbf{0.8360}} \\
				\hline 
				DNSR & \textbf{33.59}/0.9081 & 27.88/0.7674 & 26.87/0.7350 & 29.52/0.8821 & 25.72/0.7822 & 29.97/0.8331 \\
				w/ DE & 33.56/\textbf{0.9087} & \textbf{27.89}/\textcolor{red}{\textbf{0.7684}} & \textcolor{red}{\textbf{26.88}}/\textcolor{red}{\textbf{0.7358}} & \textcolor{red}{\textbf{29.60}}/\textcolor{red}{\textbf{0.8846}} & \textcolor{red}{\textbf{25.79}}/\textcolor{red}{\textbf{0.7854}} & \textbf{29.99}/\textcolor{red}{\textbf{0.8342}} \\
				\hline 
				RDN & 33.50/0.9070 & 27.87/0.7669 & 26.88/0.7357 & 29.61/0.8815 & 25.85/0.7860 & 30.05/0.8344 \\
				w/ DE & \textbf{33.51}/\textbf{0.9076} & \textcolor{red}{\textbf{27.92}}/\textcolor{red}{\textbf{0.7687}} & \textcolor{red}{\textbf{26.91}}/\textcolor{red}{\textbf{0.7367}} & \textcolor{red}{\textbf{29.67}}/\textcolor{red}{\textbf{0.8834}} & \textcolor{red}{\textbf{25.95}}/\textcolor{red}{\textbf{0.7898}} & \textbf{30.07}/\textbf{0.8352} \\
				\hline 
			\end{tabular}
		\end{small}
		\label{test_data_table}
	\end{center}
\end{table*} 

\section{Application to Scientific Datasets} \label{sds_section}

\begin{figure}
	\begin{center}
		\includegraphics[width=1\linewidth]{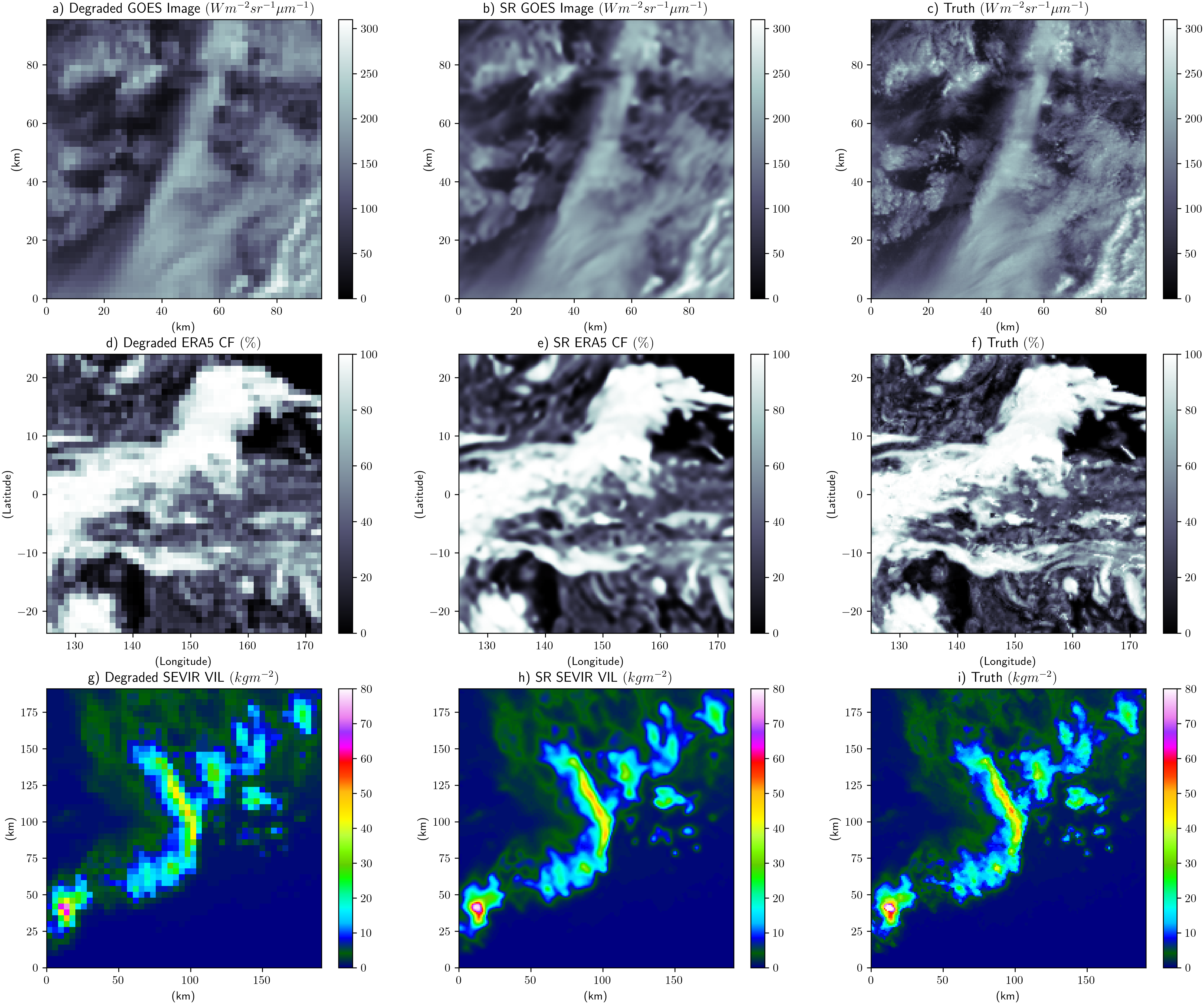}
	\end{center}
	\caption{Example degraded inputs (left), super-resolved outputs (center), and ground truth (right) for three different scientific datasets.}
	\label{sds_samples}
\end{figure}

\begin{figure}
	\begin{center}
		\includegraphics[width=1\linewidth]{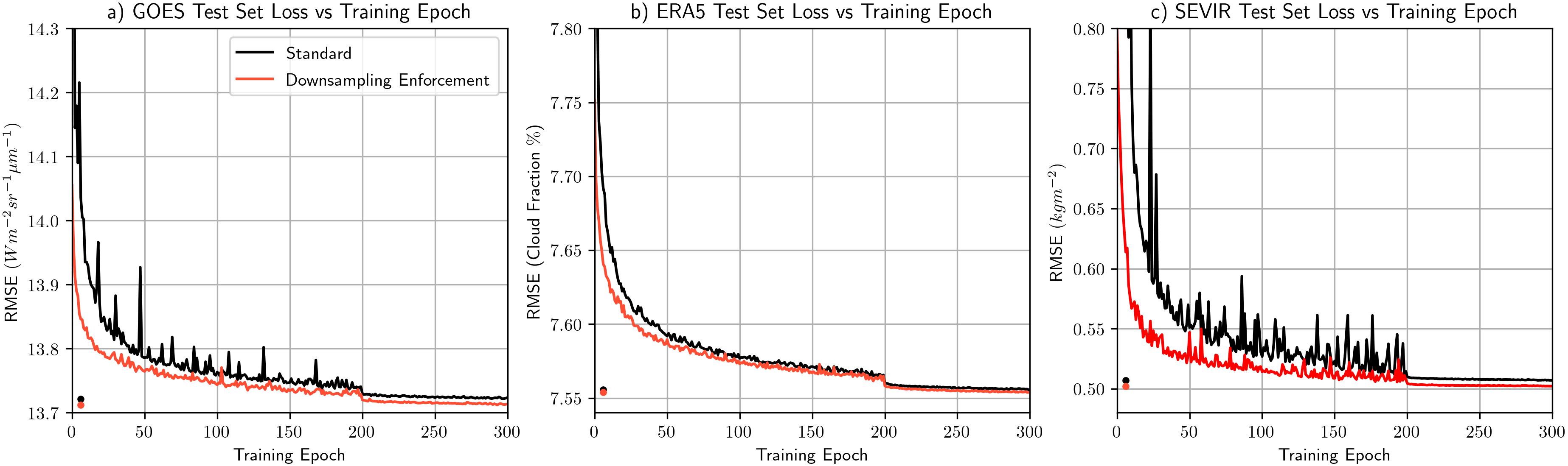}
	\end{center}
	\caption{Test set RMSE evaluated for each scientific dataset throughout training. Final test set scores (without/with DE) were: (GOES: 13.72/13.71 $Wm^{-2} sr^{-1} \mu m^{-1}$), (ERA 5: 7.56/7.55 $\%$), (SEVIR: 0.507/0.502 $kgm^{-2}$)}
	\label{sds_scores}
\end{figure}

Here we apply the DE super resolution method to three scientific datasets where strict adherence to conservation principles is important. These datasets are from diverse sources: a satellite imager, a weather radar, and a numerical weather model. In each case we use the EDRN CNN as described in Supplement Section 3 with the same training procedure described in Section \ref{training_section} (except that a lower initial learning rate of $2\times 10^{-5}$ was used for the SEVIR data). Figure \ref{sds_samples} shows an example input, degraded image, and SR output for each dataset. Figure \ref{sds_scores} shows the loss on the test sets while training (no parameter tuning was done on these datasets so no validation sets were used). In each case, inclusion of DE allows for exact enforcement of conservation laws while providing a modest improvement in performance.

\subsection{GOES-17 L1b Radiance}
The Geostationary Operational Environmental Satellite 17 (GOES-17) \cite{goes_abi_tbds} is a geostationary satellite currently orbiting above the Equatorial Pacific at 137.2W. Here we apply super resolution to level-1b radiance data from the Advanced Baseline Imager band 2 (the red 640nm band) which has a resolution of 0.5km at nadir \cite{goes_users_guide}. The images selected for this study are full-disk scans taken near 12:00 LST on various days during 2019 and 2020. The images are cropped to pixels 2452-19348 height and 8100-17700 width to avoid extreme viewing angles and low illumination near the edge of the Earth's disk. The exact file names are given in the supplement (Section 5). The L1b radiances have units of $Wm^{-2}sr^{-1}\mu m^{-1}$, and enforcing strict conservation yields a slight performance improvement for the SR scheme (Figure \ref{sds_scores}a). Example outputs are shown in Figure \ref{sds_samples} panels a-c.

\subsection{ERA5 Cloud Fraction}
The European Center for Medium Range Weather Forecasting Reanalysis version 5 (ERA5) is a reconstruction of the past state of the atmosphere from 1979-present. The reanalysis is performed by assimilating historical atmospheric observations with a numerical weather model \cite{era5}. Here we apply super-resolution to cloud fraction data which represent the fraction (0-100$\%$) of each model grid cell area covered by cloud. We use daily $0.25^{\circ} \times 0.25^{\circ}$ resolution data, between latitudes $\pm 45^{\circ}$, at 0Z for the period 1-Jan-1979 to 31-Dec-2018 for training and data from 2019 for testing. Note that these data are on a lat-lon grid while our DE technique assumes an equal area grid, so here we have used only data near the equator where there is less distortion. It is possible to modify the DE technique to include latitude weightings but we leave this for future work. Nonetheless, enforcing strict conservation rules in this context provides a performance improvement for the SR scheme (Figure \ref{sds_scores}b). Examples are shown in Figure \ref{sds_samples} panels d-f. 

\subsection{NEXRAD Vertically Integrated Liquid Water}
The Storm EVent ImageRy (SEVIR) weather dataset \cite{sevir} is composed of co-located satellite and radar observations of 20,000 weather events over the continental United States between 2017 and 2020. Here we train EDRN to perform 4x super resolution on $192\times192$ pixel (1km resolution) chips of vertically integrated liquid water (VIL), a radar derived product from the NEXRAD radar network. VIL has units of $kgm^{-2}$ and 2D average downsampling enforces conservation of mass. We use the 25th time-sample from the first 18,000 events to train and remaining 2,000 to test. Figure \ref{sds_samples} Panels g-i show an example case from the test set. The downsampling enforcement method is able to conserve liquid water mass while achieving a slight performance improvement over conventional training (Figure \ref{sds_scores}c).

\section{Additional Experiments and Discussion}
In this section we provide discussion of the limitations of our method, comparison to recent literature, and outline potential areas of future research.

\subsection{Related Methods}

No existing methods strictly enforce invertibility of CNN SR but, past studies \cite{pulse, dsloss1, dsloss2,zhang_2020} have highlighted its importance and have used loss functions to approximate invertibility. PulseGAN \cite{pulse}, adds an MSE term computed between the LR input and the downsampled output to an SR GAN's adversarial loss function, ensuring more physically plausible outputs. This method does not improve CNNs that optimize pixel-wise MSE because it effectively imposes the same loss function twice at different resolutions. We confirm this by training the EDRN-CNN with the loss function:

\begin{equation}
\mathcal{L} = MSE + \lambda \overline{(D\{x\}-D\{\hat{x}\})^2}
\end{equation}

where $MSE$ is the pixel-wise mean squared error, $x$ and $\hat{x}$ are ground truth and predicted HR pixel values respectively, $D\{*\}$ is a $4 \times 4$ averaging downsampling operator, $\lambda$ is a weighting coefficient (here, $\lambda=16$), and the over-bar represents averaging. Validation PSNR throughout training is shown in Figure \ref{training_loss}e as a blue line. The PSNR/SSIM computed on the Div2k test set was: 29.95/0.8327, lower than the scores in Table \ref{test_data_table}. The key difference of the DE approach is that it directly modifies the CNN not the loss function, and it guarantees exact instead of approximate invertibility.

\subsection{Other Downsampling Schemes} \label{bicub}

Assuming 2D-average downsampling is often correct for enforcing conservation laws, but it is not always used to train SISR CNNs. We demonstrate the impact of an incorrectly assumed downsampling scheme by training the EDRN CNN using a bicubic downsampling scheme, both with and without DE that assumes 2D-averaging. Predictably, the conventional EDRN outperforms the DE version in this case (Figure \ref{training_loss}h). For the div2k test set the conventional scheme had a PSNR/SSIM of: 30.26/0.8393 while the DE-EDRN scored: 30.13/0.8356. While 2D-averaging is used here, it may be possible to derive similar operators for other downsampling schemes, by modifying (\ref{f_piecewise} and \ref{f_compact}) to include the weights of a bicubic downsampling kernel for instance.

Finally, some of the outputs from the DE CNNs contain faint checkerboard artifacts; in the trees in the BSDS100 sample image in Supplementary Figures 1 and 2 for instance. Some checkerboarding also occurs in the no-DE cases, so we hypothesize that these patterns are partially a result of using 2D-average downsampling without anti-aliasing. In preliminary experiments we have observed that the problem is more pronounced when the DE operator is used for larger upsampling ratios (x16 upsampling) however. This is a limitation of our algorithm for large resolution increases, but an in-depth exploration of the problem and possible mitigation strategies is left as future research.

\subsection{The Magnitude of Pixel Corrections} \label{alpha_sec}

\begin{figure}
	\begin{center}
		\includegraphics[width=0.5\linewidth]{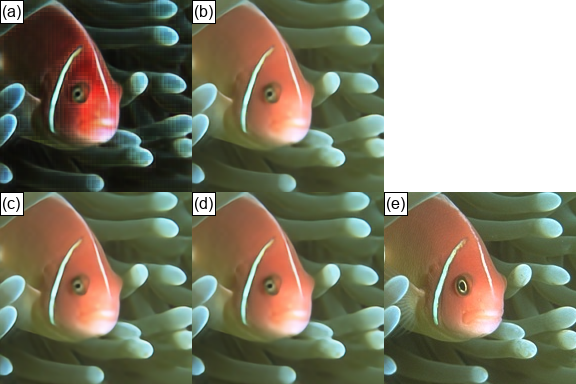}
	\end{center}
	\caption{Examination of intermediate outputs from our downsampling enforcement implementation of EDRN \cite{edrn} for a sample image from BSDS100 \cite{bsds100} before and after the DE-operator is applied. (a): output prior to DE layer; (b): final output after DE layer; (c) and (d): same as (a) and (b) but with regularization on the magnitude of the correction; (e): ground truth image for reference.}
	\label{intermediate_outputs_1}
\end{figure}

In Section \ref{f_interp_sec}, $f(\mathbf{x},P)$ is interpreted as a correction applied to an intermediate image output by the CNN. Here, we investigate whether the quality of this intermediate output improves during training, or if the CNN learns to rely on the correction by examing the value of: $|\mathbf{x} - f(\mathbf{x},P)|$, the difference between the initial output and the corrected output. After training on the Div2k test set, the average difference was 108, meaning the HR image requires a correction to ensure that it will downsample to the input image. Figure \ref{intermediate_outputs_1} panels a and b, show the intermediate output and the corrected output for an example HR image chip from the Div2k test set respectively (the original image is shown in panel e). The magnitude of the correction can be reduced by re-training with a regularizer in the loss function:

\begin{equation}
\mathcal{L} = MSE  + \lambda \overline{|\mathbf{x} - f(\mathbf{x},P)|} 
\end{equation}

where $\lambda=100$. After training with this regularization term, $|\mathbf{x} - f(\mathbf{x},P)|$ averaged over the Div2k test set was 0.3. This is demonstrated in panels c and d of Figure \ref{intermediate_outputs_1}, where the intermediate output (c) is now a near perfect match for the final output (d). Finally, the overall performance of the SISR scheme was not substantially altered and the regularized CNN had a PSNR of 30.07 and a SSIM of 0.8359 on the Div2k test set, comparable to results without regularization. Because the final outputs are nearly identical with or without the regularizer, it is not necessary to include it for most SR use cases, but the ability to increase the accuracy of the CNN's intermediate output may be useful for future applications.

\section{Conclusions}

Here, we demonstrated a new method to ensure that the output from any super-resolution-CNN, when downsampled with 2D averaging, exactly reproduces the low resolution input. In addition to providing physical consistency between the input and output data, this approach improves the CNN performance for many different super resolution architectures across several common image datasets. The method involves constructing the CNN with ``Downsampling Enforcement,'' and does not require any modifications to the data, training procedure, or loss function.

CNN-based super resolution is applicable to many types of imagery and gridded data where a guarantee that the statistics of the LR data are preserved is impactful. Here, we demonstrated how this approach could be used to: generate high resolution satellite imagery without introducing non-physical radiances; downscale coarse resolution output from a numerical model without breaking physical conservation laws; or super resolve radar data while preserving vertically integrated water mass. In these applications, preserving the LR image statistics in the HR image is paramount, and the technique presented here can deliver the high visual fidelity provided by CNN-based super resolution schemes without sacrificing physical consistency.

\newpage

\section{Supplement}

\subsection{Detailed Derivation of Equations 1 and 2}
Here, we use the notation that $\mathbf{x}$ is an $N\times N$ pixel block in the high-resolution image initially output by the CNN, where $1/N$ is the downsampling ratio. $x_i$ will represent the value of a single pixel in $\mathbf{x}$ and $P$ will represent the value of the single pixel in the low resolution input image corresponding to $\mathbf{x}$ in the output. $P$ and $x_i$ are assumed to have pixel intensities bounded by -1 and 1. $f(\mathbf{x},P)_i$ is a high-resolution pixel value after a correction $(f)$ is applied to $x_i$ (the final output of the CNN).

We also use the shorthands: $\bar{x} = \frac{1}{N} \sum_{x_i \in \mathbf{x}} x_i$, where $\bar{x}$ represents the average of the HR pixels corresponding to the single LR pixel $P$. And: $f(x_i) = f(\mathbf{x},P)_i$.

We would like $f(x_i)$ to have several properties:
\[\frac{1}{N^2} \sum_{x_i \in \mathbf{x}} f(x_i) = P \]
\[f(x_i) \in [-1,1]  \]
\[x_i > x_j \rightarrow f(x_i) \geq f(x_j) \]
\[f(x_i) \text{ is piecewise differentiable} \]
We start by considering the case that:
\[\bar{x} < P \qquad \text{where:} \quad \bar{x} = \frac{1}{N^2} \sum_{x_i \in \mathbf{x}} x_i\]
Here, the average of the pixel intensities in $\mathbf{x}$ is less than the intensity of the corresponding input pixel $P$ and the HR pixel intensities need to be adjusted upwards. $x_i \in [-1,1]$ and $f(x_i)$ can be formulated:
\[ f(x_i) = x_i + \alpha (1-x_i), \qquad \alpha \in [0,1] \]
Here, $\alpha$ is a parameter that can be solved for as a function of $P$ and $\bar{x}$.  It can be seen that when $\alpha = 1, f(x_i) = 1$, and when $\alpha=0, f(x_i) = x_i$. $\alpha$ can be found by enforcing (\ref{summation_constraint}):
\[ \alpha = (P-\bar{x})/(1-\bar{x}) \]
In the case that $\bar{x} > P$:
\[ f(x_i) = x_i - \alpha (1 + x_i), \qquad \alpha \in [0,1] \]
\[ \alpha = (\bar{x}-P)/(1+\bar{x}) \]
This adjusts the HR pixel intensities downwards so that when $\alpha=1$, $f(x_i) = -1$, and when $\alpha=0$, $f(x_i) = x_i$.\\
$f(x_i)$ can then be defined as a piecewise function:
\[
	f(x_i) = \begin{cases}
		x_i + \left(\frac{P-\bar{x}}{1-\bar{x}}\right)(1-x_i) & \bar{x} < P \\
		x_i & \bar{x} = P \\
		x_i + \left(\frac{P-\bar{x}}{1+\bar{x}}\right)(1 + x_i) & \bar{x} > P\\
	\end{cases}
\]
And finally, condensed to a single line:
\[
	f(x_i) = x_i + (P-\bar{x})\left(\frac{\sigma + x_i}{\sigma + \bar{x}}\right), \; \sigma = \text{sign}(\bar{x}-P) 
\]

\subsection{Proofs of Equations 3-6 in the manuscript}

\begin{center}\textsc{Proof of (3):}\end{center}
Given Eqn. \ref{f_compact}:
\[f(x_i) = x_i + (P-\bar{x})\left(\frac{\sigma + x_i}{\sigma + \bar{x}}\right), \quad \text{where:} \quad \sigma = \text{sgn}(\bar{x}-P), \quad \text{and:} \quad \bar{x} = \frac{1}{N^2} \sum_{x_i \in \mathbf{x}} x_i\]
\[\frac{1}{N^2} \sum_{x_i \in \mathbf{x}} f(x_i) = \frac{1}{N^2} \sum_{x_i \in \mathbf{x}} \left[x_i + (P-\bar{x})\left(\frac{\sigma + x_i}{\sigma + \bar{x}}\right)\right]\]
\[ = \frac{1}{N^2} \sum_{x_i \in \mathbf{x}} x_i + \left(\frac{P-\bar{x}}{\sigma + \bar{x}}\right) \left[ \frac{1}{N^2}  \sum_{x_i \in \mathbf{x}} \sigma + \frac{1}{N^2}  \sum_{x_i \in \mathbf{x}} x_i \right]\]
\[ =  \bar{x} + \frac{(P-\bar{x})}{(\sigma + \bar{x})} (\sigma + \bar{x}) =  \bar{x} + P-\bar{x} = P\]
\\ \\

\begin{center}\textsc{Proof of (4):}\end{center}
The remaining proofs require the piecewise definition on $f(x_i)$ given in (\ref{f_piecewise}):
\[f(x_i) = \begin{cases}
	x_i + \left(\frac{P-\bar{x}}{1-\bar{x}}\right)(1-x_i) & \bar{x} < P \\
	x_i & \bar{x} = P \\
	x_i + \left(\frac{P-\bar{x}}{1+\bar{x}}\right)(1 + x_i) & \bar{x} > P\\
\end{cases}\]
Additionally the constraint that $P, x_i, \bar{x} \in [-1,1]$ is needed.\\

\noindent \textbf{For the $\bar{x}<P$ case:} defining $\alpha = (P-\bar{x})/(1-\bar{x})$ then if $\alpha \in [0,1]$, $f(x_i) \in [x_i,1] \in [-1,1]$. \\
\[\text{Starting from:} \quad  \bar{x} < P \leq 1\]
\[0 < P-\bar{x} \leq 1-\bar{x}\]
\[\text{noting that:} \quad \bar{x} < 1 \rightarrow -\bar{x} > -1 \rightarrow 1-\bar{x} > 0\]
\[0 < \frac{P-\bar{x}}{1-\bar{x}} \leq 1 \quad \rightarrow \alpha \in [0,1] \quad \text{and} \quad f(x_i) \in [-1,1]\]
\textbf{The $\bar{x} = P$ case} is trivial because $f(x_i) = x_i$ and since $x_i \in [-1,1]$, $f(x_i) \in [-1,1]$.\\

\noindent \textbf{For the $\bar{x}>P$ case:} defining $\alpha = (\bar{x}-P)/(\bar{x}+1)$ then $f(x_i) = x_i - \alpha (1+x_i)$ and if $\alpha \in [0,1]$, $f(x_i) \in [-1,x_i] \in [-1,1]$. \\
\[\text{Starting from:} \quad  -1 \leq P < \bar{x} \]
\[ 1 \geq -P > -\bar{x}\]
\[\bar{x} + 1 \geq \bar{x} -P > 0\]
\[\text{noting that:} \quad \bar{x} > -1 \rightarrow 1+\bar{x} > 0\]
\[1 \geq \frac{\bar{x}-P}{\bar{x}+1} > 0 \quad \rightarrow \alpha \in [0,1] \quad \text{and} \quad f(x_i) \in [-1,1]\]
\\ \\

\begin{center}\textsc{Proof of (5):}\end{center}
We would like $f(x_i)$ to have the property that $x_i > x_j \rightarrow f(x_i) \geq f(x_j)$. Note that here $x_i$ and $x_j$ are high resolution pixel value outputs from the algorithm that each correspond to the same input pixel $P$. This means that their corresponding values of $P$ and $\bar{x}$ are the same. Again consider Eqn. \ref{f_compact}:\\

\noindent\textbf{For the $\bar{x}<P$ case:} taking $\alpha = (P-\bar{x})/(1-\bar{x})$, $f(x_i) = x_i + \alpha(1-x_i)$:\\
We know from the proof of (2) that for $\bar{x} < P$: $\alpha \in [0,1]$ so $(1-\alpha) \in [0,1]$ and:
\[\text{if} \quad x_i > x_j\]
\[x_i (1-\alpha)  \geq  x_j(1 - \alpha)\]
\[x_i + \alpha - \alpha x_i \geq x_j + \alpha - \alpha x_j\]
\[x_i + \alpha(1- x_i) \geq x_j + \alpha(1 - x_j)\]
\[f(x_i) \geq f(x_j)\]

\noindent\textbf{For the $\bar{x}=P$ case:} this is trivial because $f(x_i) = x_i$ and $f(x_j) = x_j$ so if $x_i>x_j \rightarrow f(x_i)>f(x_j)$.\\

\noindent\textbf{For the $\bar{x}>P$ case:} taking $\alpha = (\bar{x}-P)/(\bar{x}+1)$, and $f(x_i) = x_i-\alpha(1+x_i)$:\\
We know from the proof of (2) that for $\bar{x} > P$: $\alpha \in [0,1]$ so $(1-\alpha) \in [0,1]$ and:
\[\text{if} \quad x_i > x_j\]
\[x_i (1-\alpha)  \geq  x_j(1 - \alpha)\]
\[x_i - \alpha - \alpha x_i \geq x_j - \alpha - \alpha x_j\]
\[x_i - \alpha(1+ x_i) \geq x_j -\alpha(1 + x_j)\]
\[f(x_i) \geq f(x_j)\]
\\

\begin{center}\textsc{Proof of (6):}\end{center}
Here, we use the piecewise definition of $f(x_i)$ given in (\ref{f_piecewise}) and demonstrate its differentiability by finding the partial derivatives with respect to $x_i$. Note that $\bar{x}$ is a function of $x_i$ so $\partial \bar{x} / \partial x_i = N^{-2}$:\\
For $\bar{x} < P$:
\[f(x_i) = x_i + \frac{P-\bar{x}}{1-\bar{x}} (1-x_i) = \frac{x_i-x_i\bar{x}}{1-\bar{x}} + \frac{P-Px_i - \bar{x} + \bar{x} x_i}{1-\bar{x}} = \frac{x_i+P-Px_i-\bar{x}}{1-\bar{x}}\]
\[\frac{\partial f}{\partial x_i} = \left[ (1-\bar{x})(1-P-N^{-2}) + N^{-2}(x_i+P-Px_i-\bar{x})\right]/(1-\bar{x})^2 \]
\[=\left( 1 - P - N^{-2} - \bar{x} + P \bar{x} + N^{-2} \bar{x} + N^{-2} x_i + N^{-2} P - N^{2} Px_i - N^{-2} \bar{x} \right) / (1-\bar{x})^2\]
\[=(P-1)(\bar{x} - 1 + (1 - x_i)/N^2)(1-\bar{x})^{-2}\]
We omit the derivation of the other derivative as it is very similar, the full derivative of $f(x_i)$ is:
\[\frac{\partial f(x_i)}{\partial x_i} = \begin{cases}
	(P-1)(\bar{x} - 1 + (1 - x_i)N^{-2})(1-\bar{x})^{-2}& \bar{x} < P \\
	1 & \bar{x} = P \\
	(P+1)(\bar{x} + 1 - (1 + x_i)N^{-2})(1+\bar{x})^{-2}& \bar{x} > P\\
\end{cases}\]

\subsection{Individual CNN Descriptions}
\subsubsection{SR-CNN}
\cite{srcnn} is an early application of convolutional neural networks to the task of image super resolution. The authors use a relatively simple CNN with only three convolutional layers and ReLU transfer functions to map a LR image to its HR counterpart. Here, we apply the so-called ``9-5-5'' version of this model. It consists of a 3-layers with 64-9x9 filters, 32-5x5 filters, and 3-5x5 filters in that order. As implemented here, it has $6.9\times 10^4$ trainable parameters. Unsurprisingly, it is outperformed by all of the other CNNs, all of which have much more sophisticated design and many more trainable parameters.
\subsubsection{Lap-SRN}
\cite{lapsrn} design a CNN based on Laplacian pyramid upsampling. While some earlier approaches \cite{srcnn} first upsampled images using an interpolation scheme, and then processed the upsampled image with a CNN, \cite{lapsrn} instead operate directly on the LR image, and  progressively upsample it using transposed convolutions. Their CNN consists of two branches, a feature extraction branch, that infers HR residual features from the LR image, and a image reconstruction branch, that generates images at the intermediate resolutions between the LR input and the HR output by adding the residuals produced by the feature extraction branch. Our implementation of the Lap-SRN scheme uses $8.4\times 10^5$ trainable parameters. It is outperformed by all of the CNNs but SR-CNN and the Dense U-Net. It is also one of the earlier implementations of CNN-based SISR.
\subsubsection{Dense U-Net (DUN)}
This CNN combines concepts from \cite{unet} and \cite{densenet}. It is a U-net style architecture where each of the 2-3 convolution blocks in the U-net have been replaced with a block of densely connected convolutions. This architecture has previously been used for applying super-resolution to weather radar data \cite{radarsr}. The U-net architecture was originally developed for image segmentation, but is particularly good at combining feature information at different spatial scales \cite{unet}, and so seems appropriate for the SISR task. The implementation used here uses a growth rate of 38 (see \cite{densenet}), downsamples the input to 1/4 of its original resolution at the lowest level of the U-net, and uses 3 convolutions per densely connected block. It has $5.3\times 10^6$ trainable parameters.
\subsubsection{Deep Back Projection Network (DBPN)}
The deep back projection network \cite{dbpn} uses a series of blocks that repeatedly upsample and then downsample the input image between the input and target resolutions using transposed and strided convolutions (respectively). The high- and low- resolution feature representations from prior blocks are concatenated and fed into subsequent blocks, much like in a densely connected network \cite{densenet}. In our implementation for 4x SR we use a filter size of 8x8 for the up- and down-sampling layers. We use a total of eight blocks where each block consists of four convolutional layers organized as either: up-down-up-down or down-up-down-up, with the order depending on the input resolution. Each of the convolutional layers use 64 filters. This results in $5.6\times 10^6$ trainable parameters.
\subsubsection{Dense SR Net (DSRN)}
This CNN directly extends concepts from \cite{densenet} to super resolution \cite{densesr}. The network consists of a series of densely connected blocks of convolutional layers. Dense blocks concatenate the feature outputs of all previous convolutional layers before feeding them into the next one. The intuition behind this is that while the number of features in each individual layer is smaller than other CNN architectures, the CNN can learn to use combinations of different features from diffrent parts of the network. It also provides more direct paths between the input and output of the network, which helps reduce the vanishing gradient problem. The blocks operate at low spatial resolution and build up a feature representation of the image. At the end of the network there is an upsampling module composed of two transposed convolutions. In our implementation, there are 8 dense blocks composed of 6 layers each with a growth rate of 16. The upsampling module uses 256 features. This results in $5.8\times 10^6$ trainable parameters.
\subsubsection{Enhanced Deep Residual Network (EDRN)}
\cite{edrn} is based on ideas from \cite{resnet}. This network is composed of a series of residual blocks. These blocks pass the input to the block through several convolutional layers and then add the result to the original input. The intuition here is that these residual skip connections provide a more direct path between the input and output of the network, and allow the neural network to learn which residual blocks and which features to use as it trains, again combating the vanishing gradient problem. ``Enhanced" refers to removing several aspects of the residual blocks that were used in \cite{resnet} that are not beneficial for super resolution. Batch normalization for instance. Our implementation has 16 residual blocks with 128-filter internal convolutional layers. Like DSRN, this approach operates at low resolution and then upsamples as one of the final operations in the network architectures. Here the upsampling is done using a pixel shuffle however \cite{pixel_shuffle}. The implementation used here has $5.2\times 10^6$ trainable parameters.
\subsubsection{Residual Dense Network (RDN)}
The residual dense network \cite{rdn}, as the name suggests, combines aspects of both residual and densely connected networks. Like the last two CNNs discussed, it operates at low spatial resolution and builds up a feature representation of the image before upsampling it in the final layers of the network. Internally, it is composed of densely connected blocks, each block however is followed by a feature compression layer (1x1 convolution) that allows the input to the block to be added to the output (a residual connection). Finally, the network includes a long skip connection between the first layers and the last. Upsampling is performed with a pixel shuffle. Our implementation uses 10 blocks of 9 layers each with 32 channels per layer. This yields $5.3 \times 10^6$ parameters.

\subsection{Example CNN Outputs}
(Next page)
\begin{figure*}
	\begin{center}
		\includegraphics[width=1\linewidth]{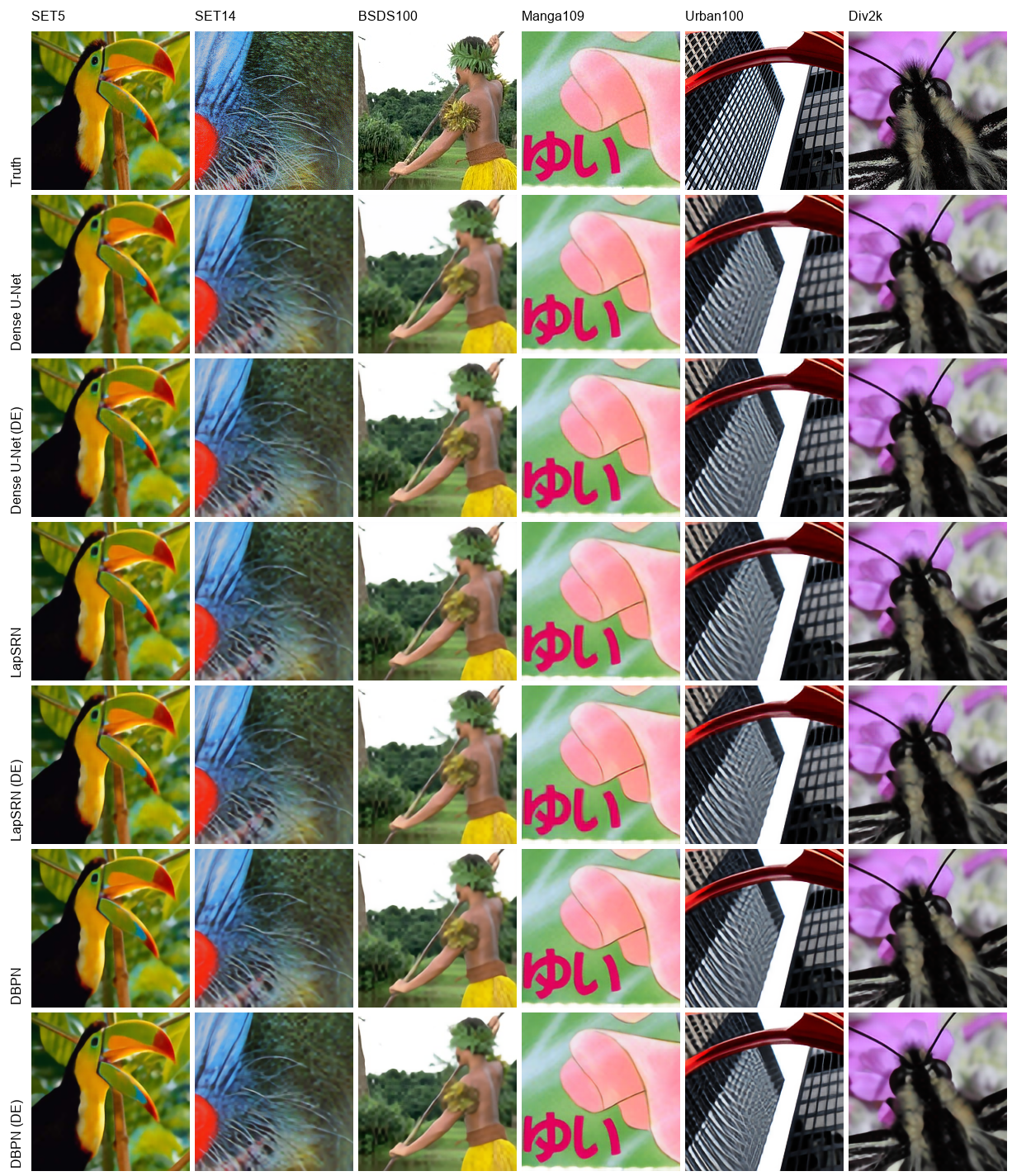}
	\end{center}
	\caption{An example of applying super resolution to an image chip from each of the training sets for three of the CNNs, both with and without downsampling enforcement.}
	\label{example_outputs1}
\end{figure*}

\begin{figure*}
	\begin{center}
		\includegraphics[width=1\linewidth]{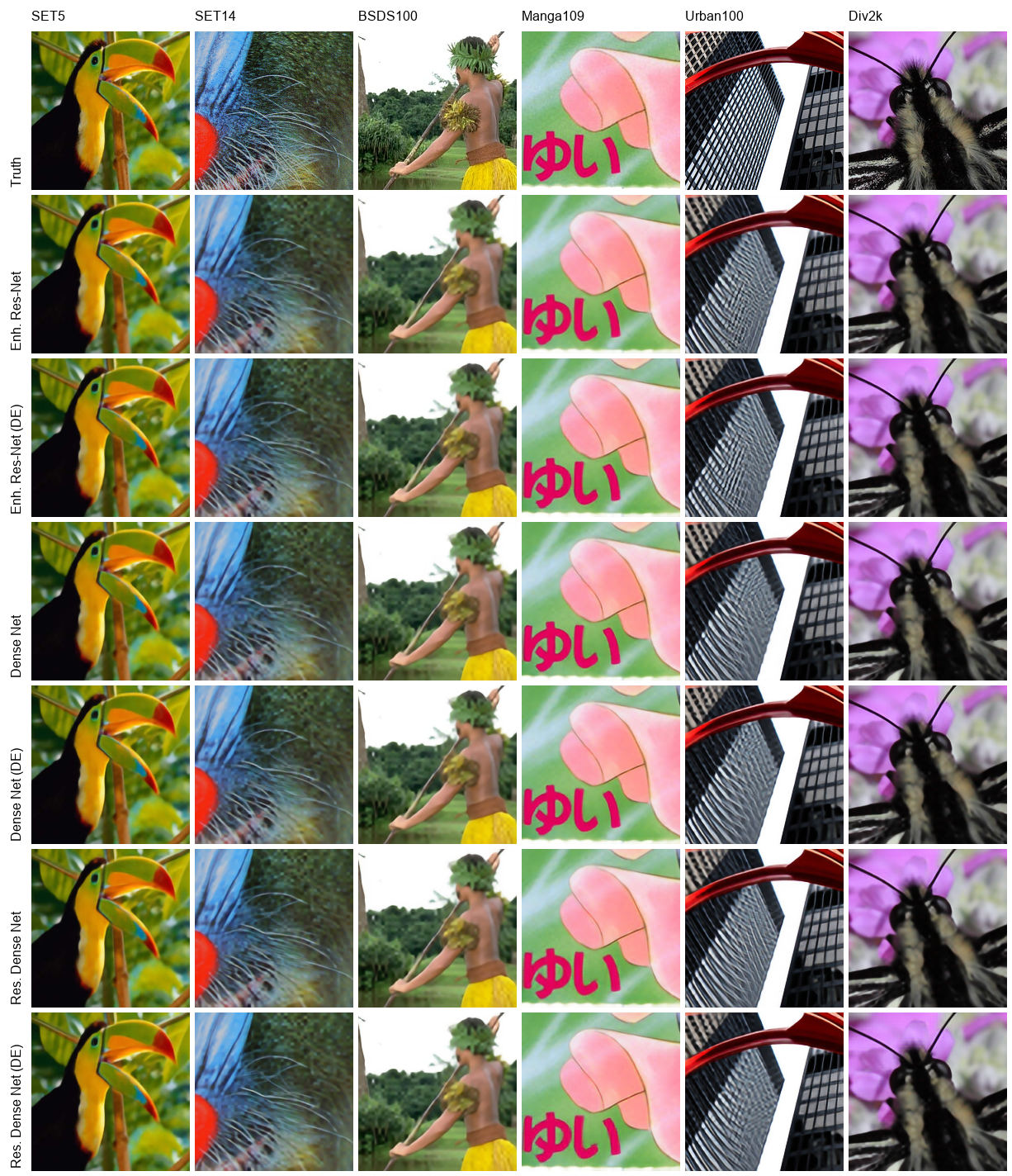}
	\end{center}
	\caption{An example of applying super resolution to an image chip from each of the training sets for three of the CNNs not shown in Figure \ref{example_outputs1}, both with and without downsampling enforcement.}
	\label{example_outputs2}
\end{figure*}

\newpage
\subsection{List of GOES Files}
\begin{center}\textsc{Training Set:}\end{center}
{\small
	\begin{verbatim}
		OR_ABI-L1b-RadF-M3C02_G17_s20200501830354_e20200501841121_c20200501841151.nc
		OR_ABI-L1b-RadF-M6C02_G17_s20200101830321_e20200101839388_c20200101839418.nc
		OR_ABI-L1b-RadF-M6C02_G17_s20200201820321_e20200201829388_c20200201829415.nc
		OR_ABI-L1b-RadF-M6C02_G17_s20200301820321_e20200301829388_c20200301829425.nc
		OR_ABI-L1b-RadF-M6C02_G17_s20200401840321_e20200401849388_c20200401849427.nc
		OR_ABI-L1b-RadF-M6C02_G17_s20200601820321_e20200601829387_c20200601829424.nc
		OR_ABI-L1b-RadF-M6C02_G17_s20200701810300_e20200701819367_c20200701819385.nc
		OR_ABI-L1b-RadF-M6C02_G17_s20200801810321_e20200801819388_c20200801819424.nc
		OR_ABI-L1b-RadF-M6C02_G17_s20200901850321_e20200901859388_c20200901859413.nc
		OR_ABI-L1b-RadF-M6C02_G17_s20201001810321_e20201001819388_c20201001819416.nc
		OR_ABI-L1b-RadF-M6C02_G17_s20201101820321_e20201101829388_c20201101829414.nc
		OR_ABI-L1b-RadF-M6C02_G17_s20201201850321_e20201201859388_c20201201859432.nc
		OR_ABI-L1b-RadF-M6C02_G17_s20201301820321_e20201301829388_c20201301829413.nc
		OR_ABI-L1b-RadF-M6C02_G17_s20201401840321_e20201401849388_c20201401849419.nc
		OR_ABI-L1b-RadF-M6C02_G17_s20201501810321_e20201501819388_c20201501819425.nc
		OR_ABI-L1b-RadF-M6C02_G17_s20201601820321_e20201601829388_c20201601829412.nc
		OR_ABI-L1b-RadF-M6C02_G17_s20201701840321_e20201701849388_c20201701849418.nc
		OR_ABI-L1b-RadF-M6C02_G17_s20201801850321_e20201801859388_c20201801859428.nc
		OR_ABI-L1b-RadF-M6C02_G17_s20201901810321_e20201901819388_c20201901819426.nc
		OR_ABI-L1b-RadF-M6C02_G17_s20202001820321_e20202001829388_c20202001829413.nc
		OR_ABI-L1b-RadF-M6C02_G17_s20202101830321_e20202101839388_c20202101839416.nc
		OR_ABI-L1b-RadF-M6C02_G17_s20202201800321_e20202201809388_c20202201809415.nc
		OR_ABI-L1b-RadF-M6C02_G17_s20202301840321_e20202301849388_c20202301849425.nc
		OR_ABI-L1b-RadF-M6C02_G17_s20202401800321_e20202401809388_c20202401809410.nc
		OR_ABI-L1b-RadF-M6C02_G17_s20202501820321_e20202501829388_c20202501829424.nc
		OR_ABI-L1b-RadF-M6C02_G17_s20202601800321_e20202601809388_c20202601809425.nc
		OR_ABI-L1b-RadF-M6C02_G17_s20202701850321_e20202701859388_c20202701859413.nc
		OR_ABI-L1b-RadF-M6C02_G17_s20202801800321_e20202801809387_c20202801809423.nc
		OR_ABI-L1b-RadF-M6C02_G17_s20202901810321_e20202901819387_c20202901819415.nc
		OR_ABI-L1b-RadF-M6C02_G17_s20203001830321_e20203001839388_c20203001839422.nc
		OR_ABI-L1b-RadF-M6C02_G17_s20203101830321_e20203101839388_c20203101839425.nc
		OR_ABI-L1b-RadF-M6C02_G17_s20203201810321_e20203201819388_c20203201819410.nc
		OR_ABI-L1b-RadF-M6C02_G17_s20203301800321_e20203301809388_c20203301809426.nc
		OR_ABI-L1b-RadF-M6C02_G17_s20203401830321_e20203401839388_c20203401839424.nc
		OR_ABI-L1b-RadF-M6C02_G17_s20203501850321_e20203501859388_c20203501859422.nc
		OR_ABI-L1b-RadF-M6C02_G17_s20203601800321_e20203601809388_c20203601809414.nc
\end{verbatim}}

\begin{center}\textsc{Test Set:}\end{center}
{\small
	\begin{verbatim}
		OR_ABI-L1b-RadF-M6C02_G17_s20191001800341_e20191001809408_c20191001809435.nc
		OR_ABI-L1b-RadF-M6C02_G17_s20192001810341_e20192001819408_c20192001819432.nc
		OR_ABI-L1b-RadF-M6C02_G17_s20193001840341_e20193001849407_c20193001849432.nc
\end{verbatim}}

\end{document}